# The effect of tow gaps on compression after impact strength of robotically laminated structures

A. T. Rhead[1], T. J. Dodwell[1] and R. Butler[1]*

**Abstract** When (robotic) Automated Fibre Placement (AFP) is used to manufacture aerospace components with complex three dimensional geometries, gaps between fibre tows can occur. This paper is the first to explore the interaction under compressive load of these tow gaps with impact damage. Two coupons with different distributions of tow-gaps were impacted. Results indicated that the area of delamination is smaller for an impact directly over a tow gap where the tow gap is situated close to the non-impact face. Subsequent Compression After Impact (CAI) testing demonstrated that both the formation of sublaminate buckles and subsequent growth of delaminations is inhibited by the presence of a tow gap near the non-impact face. Non-destructive testing techniques and a computationally efficient infinite Strip model are used to analyse the damage resistance and damage tolerance of the coupons. A new validation of the Strip model is also presented.

## 1 Introduction

Automated fibre placement (AFP) technologies [Gurdal et al. (2008); Gurdal and Olmedo (1993); Croft et al. (2011)] allow the rapid production of Carbon Fibre Reinforced Plastic (CFRP) structures. AFP and more advanced techniques such Continuous Tow Shearing (CTS) [Kim et al. (2012)] are also allowing fibres to be steered in the plane of the laminate, allowing improvements in buckling and laminate stiffness to be realised [Liu, Butler (2013); Raju et al. (2013)]. However, unlike CTS, steering of fibres with AFP can result in either gaps between tows or overlaps of tows. This is because, in current AFP processes, the tow placement head is constrained to remain perpendicular to the direction in which tows are laid [Fayazbakhsh et al. (2012)]; thus ensuring each course of tows maintains a constant ply width. This allows courses of straight fibre tows to be laid down efficiently on flat or gently curved surfaces. However, in parts with more complex geometries, the restraint of the fibre placement head can cause gaps or overlaps to form between courses. An illustrative example of the formation of tow gaps in a Tapered Channel Section (TCS) is described below.

A TCS is representative of the principal load carrying structures that run from the root to the tip of an aircraft wing. As can be seen from Fig. 1(a) a TCS is comprised of two flanges (F1 and F2) and a web (W). A TCS is manufactured by building up courses of AFP tows over a mandrel tool. Each course of AFP tows is typically aligned at 0°, ±45° or 90° to the principal loading direction. The cross-section of the TCS changes along its length as a result of a taper in the wing, Fig. 1(b), meaning flange F2 lies at an angle of θ° to flange F1. As each flange is predominately loaded in shear, the precise alignment of 45° fibres in these regions is important. If an AFP machine started by laying a straight 45° course of tows across F1, the course of tows would remain at 45° on web W, yet would be at 45 + θ° on flange F2. In practice, with restriction of the AFP head in mind, the approach taken is to sacrifice the fibre alignment over the web by steering tows through an angle θ° across W. This ensures that both flanges F1 and F2 will have fibres at 45°. To understand this first consider two courses, each of 8 tows, being laid adjacent to one another without gaps. The constraint that the normal width of the layers must remain constant, coupled with no gaps being allowed between courses, forces the curvature of consecutive tows to tighten, quickly leading to a non-manufacturable *geometric singularity* [Dodwell et al. (2012)], see Fig. 2(a). Even before this singularity, there is a limit to the curvature which an AFP head can produce. High curvatures lead to extra compression being induced on the inner side of the course, making tows susceptible to micro-buckling and other defects [Beakou et al. (2011)]. The current solution is a compromise of laying identical courses of curved tows adjacent to one another but allowing, where necessary, gaps to form between each course, as shown in Fig. 2(b).

In-service, components containing tow gaps, like all other composite airframe structures, would be susceptible to impact damage. Therefore the study of the effect of such gaps on impact damage and post impact performance is vital. Impact of aircraft components falls into two categories, damage resistance (how much damage is incurred for a given impact) and damage tolerance or Compression After Impact (CAI) strength (the amount of compressive strain the damaged structure can tolerate before failure), both of which are interlinked and studied in this paper. A particularly dangerous form of damage is Barely Visible Impact Damage (BVID). Damage is characterised as BVID if it causes surface deformations just below the limit of detectability on a standard, in-service, visual inspection of an aircraft. BVID mainly comprises intra-ply cracks and delaminations (separations of plies) of which, under compressive loading, the latter may propagate, ultimately causing failure of the component. Currently, failure is prevented by setting a damage tolerance strain allowable for the component below the strain required to cause delamination propagation. However, if any reduction were required in the damage tolerant strain allowable due to the presence of tow gaps, the structure would need to be thickened. This would result in an increase in aircraft weight at the cost of fuel efficiency.

[1]Composites Research Unit Department of Mechanical Engineering, University of Bath, Claverton Down, Bath, BA2 7AY, UK
*R.Butler@bath.ac.uk

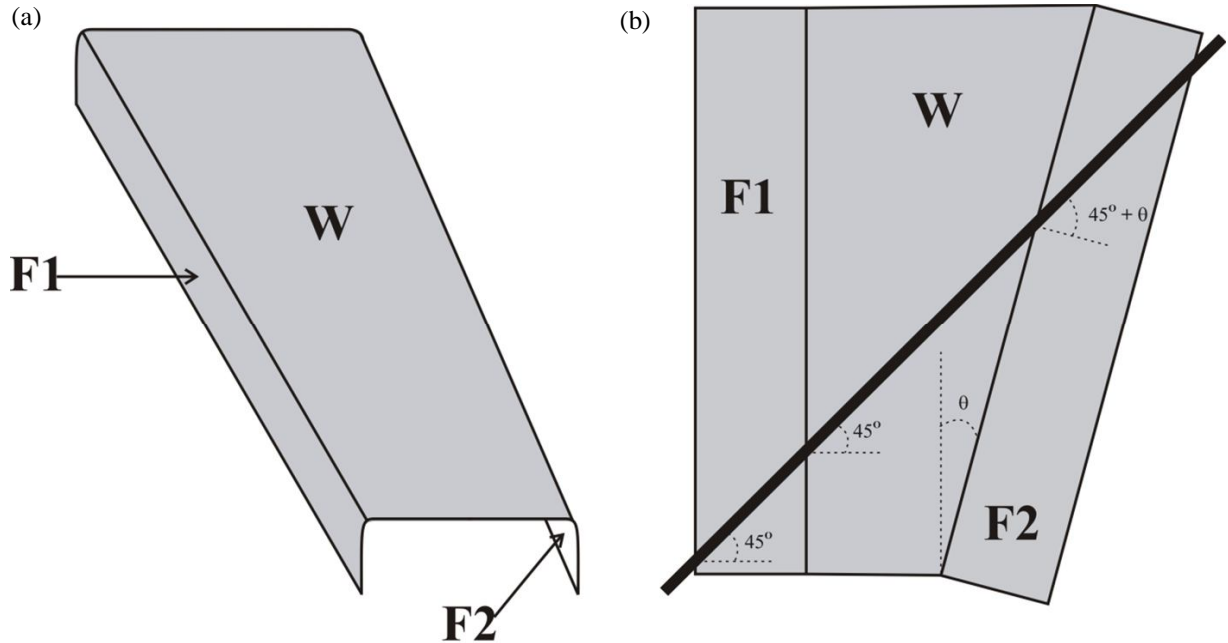

**Figure 1** (a) Shows a typical configuration of a TCS, note the tapering cross-section. (b) Shows an unfolded view of the channel demonstrating the path of a 45° fibre. Note that the path does not remain at 45° over surface F2.

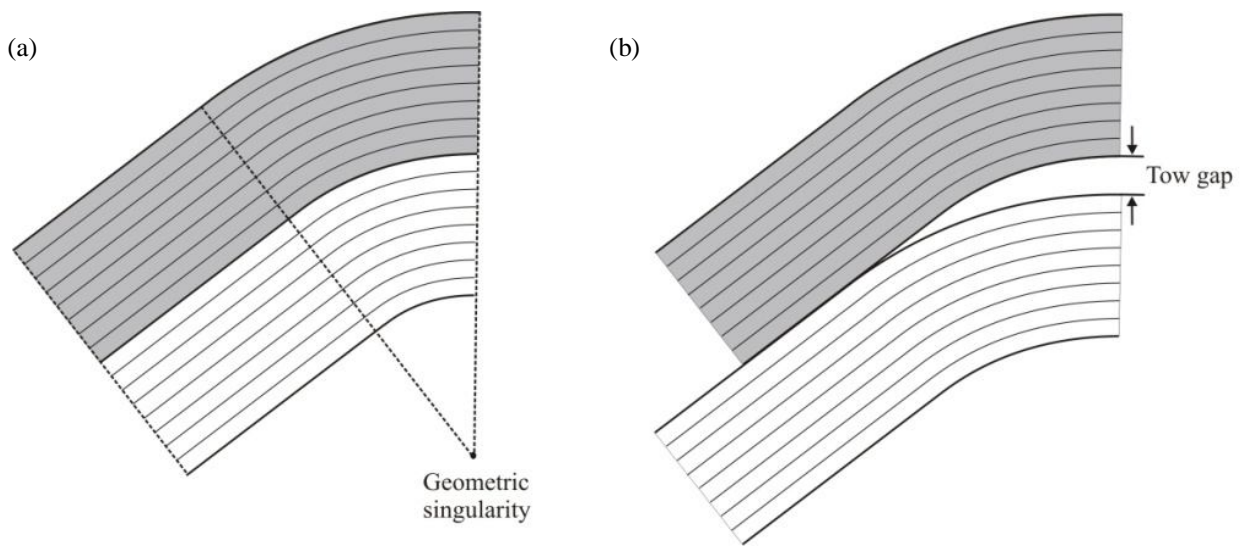

**Figure 2** Two possible configurations for adjacent courses of 8 steered tows. (a) Courses are laid down so that no gaps form between each course, consequently the curvature of each consecutive tow is forced to tighten. (b) By laying down identical courses next to one another, no such tightening of curvatures occurs, but at the sacrifice of gaps between courses.

This paper will present and discuss the first ever results of tests aimed at understanding the effect of tow gap distribution on both damage morphology and CAI strength. A Strip model [Butler et al. (2012)] for sublaminate buckling-driven propagation of delaminations developed by the authors will be used to aid analysis of experimental results. A new validation of this Strip model using tests on artificially delaminated laminates taken from the literature will also be presented. It will be shown that impact of a coupon directly over a tow gap that lies close to the non-impact surface (the through-thickness region which contains the delaminations that are likely to propagate under compressive loading) produces a smaller damage area than impact to a region with a tow gap near the impact surface. Results of CAI tests will also show that the presence of tow gaps near the non-impact surface can inhibit sublaminate buckling and growth of delaminations, the most critical mechanism for CAI failure.

**2 Strip model for sublaminate buckling and delamination propagation**

The Strip model predicts critical threshold values of compressive axial strain below which local sublaminate buckling-driven propagation of delaminations will not occur. It is assumed that the boundaries of the delaminations are circular or can be approximated by a circle [Rhead and Butler (2009)]. The Strip model is an equivalent model that does not represent exact physical reality. Instead it seeks to release the equivalent value of elastic energy stored in the post-buckled sublaminate in pure Mode I fracture (peeling); a simplification of the mixed mode conditions detected in the full 3D reality. A comparison of bending and membrane energies in the sublaminate prior to and following propagation is used to derive an equation for the threshold strain, $\varepsilon_{th}$, the strain below which delamination propagation will not occur,

$$\varepsilon_{th} = \varepsilon^C \left( \sqrt{4 + \frac{2G_{IC}}{(\varepsilon^C)^2 A_{11}}} - 1 \right) \tag{1}$$

Here $A_{11}$ is the axial stiffness of the sublaminate, $G_{IC}$ is the strain energy release rate required to cause Mode I failure of the matrix and $\varepsilon^C$ is the sublaminate buckling strain calculated using the infinite strip program VICONOPT [Williams et al. (1991)]. Other methods for calculating the buckling strain such as Finite Element Analysis (FEA) can be used, however this is likely to result in a considerable loss in computational efficiency in comparison with VICONOPT. A useful description of the Strip model using an equivalent sandwich strut analogy is given in [Rhead et al. (2012)] and a full derivation can be found in [Butler et al. (2012)]. Note that the number of plies making up the sublaminate is assumed to remain constant.

The Strip model has previously been validated against both CAI tests [Rhead et al. (2011)] and compression tests on artificially delaminated laminates [Butler et al. (2012); Rhead et al. (2012)]. Here, further validation is achieved by comparison with results from Reeder et al. (2002) for compression testing of artificially delaminated coupons manufactured from an AS4/3501-6 graphite/epoxy CFRP material with stacking sequence [(-45/45/90/0)$_2$/-60/60/-15/15]$_S$. Material properties can be found in Reeder et al. (2002) except for $G_{IC}$ = 128 J/m$^2$ [Hexcel, (1998)]. Artificial delaminations were created using circular Teflon films 64mm in diameter placed at either the 4$^{th}$ or 5$^{th}$ layer interface. In Tab. 1 averaged results for $\varepsilon^C$ and $\varepsilon_{th}$ from these compression tests are compared with results from the Strip model and FEA employing the Virtual Crack Closure Technique (VCCT) [Reeder et al. (2002)]. Experimental results for $\varepsilon^C$ are inflated owing to residual adhesion between the Teflon insert and the sublaminate. Strip model results for $\varepsilon_{th}$ given in Tab. 1 are conservative and accurate to within 14% of experimental results and 7% of FEA.

**Table 1**. Comparison of Strip model results for sublaminate buckling and delamination propagation strains with experimental and FEA results from Reeder et al. (2002).

| Variable | Delaminated interface | Experimental (μstrain) | FEA (μstrain) | Strip model (μstrain) |
|---|---|---|---|---|
| $\varepsilon^C$ | 4 | 2250 | 700 | 628 |
| $\varepsilon_{th}$ | 4 | 2600 | 2700 | 2532 |
| $\varepsilon^C$ | 5 | 2400 | 850 | 971 |
| $\varepsilon_{th}$ | 5 | 2700 | 2500 | 2322 |

**3 Materials and test methods for specimens with tow gaps**

Two coupons were cut from an area of a TCS with stacking sequence [±45/0/-45/90/0$_2$/45/0$_2$/±45/0$_2$/45/0$_2$/90/-45/0/∓45]. The TCS was manufactured from 0.25mm thick Hexcel M21/IMA-12k pre-preg CFRP tows with material properties $E_{11}$ = 145GPa, $E_{22}$ = 8.5GPa, $G_{12}$ = 4.2GPa, $v_{12}$ = 0.35 and $G_{IC}$ = 500J/m$^2$. The Coriolis Composites AFP machine used to manufacture the TCS lays courses of up to 8 tows at a time with a total width of 50.1mm per course. Coupon dimensions and a diagram of areal tow gap positions (as viewed from the free (non-tool) surface) for both coupons A and B are given on the right hand side of Fig. 3.

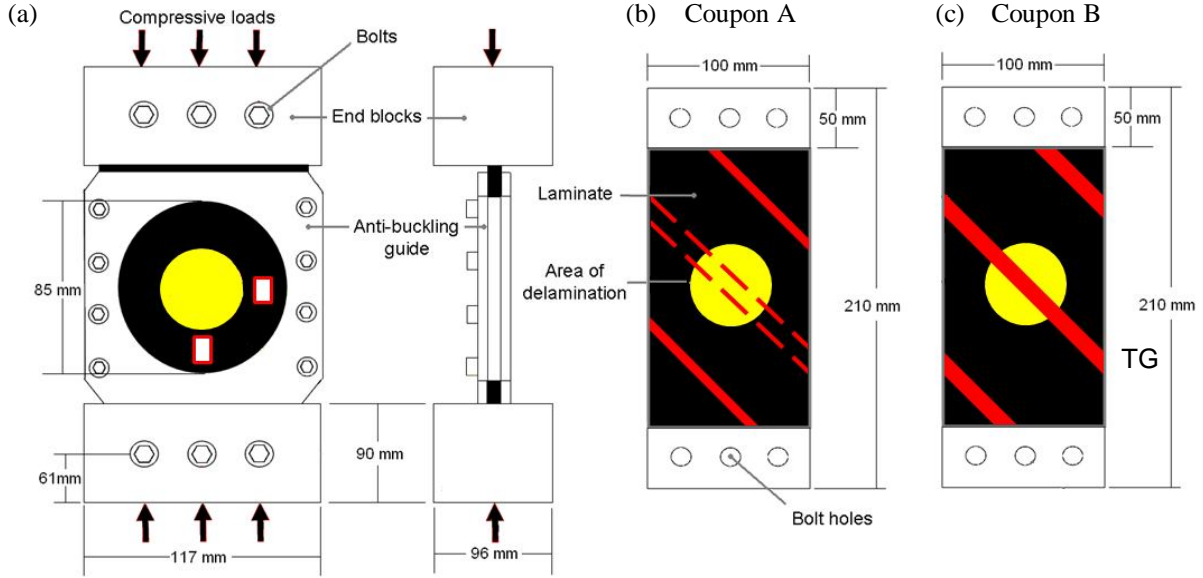

**Figure 3.** (a) Compression after impact test fixture, (b) and (c) schematics of coupons A and B respectively showing areal positioning of tow gaps and impact sites on the free surface. TG indicates the position of the photographic cross-section in Fig. 4(a).

A photographic cross-section of a tow gap in the 8$^{th}$ (45°) ply of coupon B together with surface images, taken using an Ultrasonic Sciences Ltd C-scan system, caused by tow gaps in the vicinity of the impact point (central to the image) can be seen in Fig. 4. As a consequence of the consolidation of plies in the manufacturing process, tow gaps manifest themselves as approximately 3mm wide and 0.25-0.5mm deep channels on the free (non-tool) surface of coupon B. The tow gap seen in the centre of Fig. 4(c) occurs in the 14$^{th}$ ply. The distance of this tow gap from the free surface means little surface distortion is present in comparison to coupon B. This can be seen as a comparative lack of definition of the gap in Fig. 4(b). Further channels are seen in coupon A but occur in the first ply and hence are present on the free surface only. No channels are present on the tool surface of either coupon.

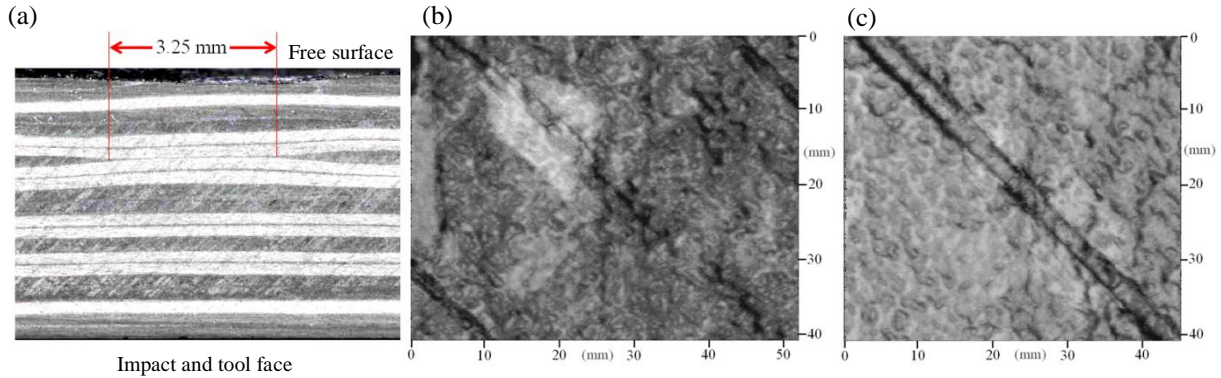

**Figure 4.** Tow gaps images: (a) cross-section photograph of the edge of coupon B at point TG in Fig. 3(c) showing a tow gap in the 8$^{th}$ layer. (b) and (c) C-scan surface images of coupons A and B respectively.

Coupons were subject to 18J impacts at the plan form centre of their flat tool face. Impacts were delivered by an Instron Dynatup 9250 HV instrumented drop weight impact machine employing a 16mm hemispherical tup. Coupons were held during impact across a 125 mm by 75 mm window as per ASTM standard D7137 [ASTM D7136 (2009)]. Coupons were C-scanned following impact to establish the induced damage morphology.

Following impact coupons were axially compressed until failure in an Instron 5585H compression test machine at a displacement rate of 0.1 mm/min. During compression, coupons were restrained against overall buckling by an anti-buckling guide see Fig. 3(a). A Digital Image Correlation (DIC) system, employing a pair of stereo cameras was used to measure the 3D surface displacement of the laminates in relation to their unloaded position. This allowed the visualization of buckling modes and delamination growth following post-processing. To ensure panels were correctly aligned and placed under pure axial compression, strains were recorded throughout the tests by two pairs of vertically aligned back-to-back strain gauges, see Fig. 3(a). A comparison of CAI strength with a coupon completely devoid of tow gaps was not undertaken. The presence of tow gaps in tapered components manufactured using AFP is unavoidable (as is shown in Fig.1), and hence such a comparison would be unrepresentative of the application.

## 4 Results for tow gap specimens

*4.1 Impact results*

Coupons were nominally impacted at 18J. However a comparison of impact plots for coupons A and B in Fig. 5 shows that the peak energy received was 18.1 J and 18.6 J for coupons A and B respectively. Areas under the energy curves in Fig. 5 and peak deflection data indicates that coupon B exhibited a more elastic response to impact. Peak impact loads were 9.6 kN and 9.4 kN for coupons A and B respectively. Post-impact C-scan images are shown in Fig. 6. Figures 6 (a) and (b) show surface damage to the tool surface following impact and Figs. 6 (c) and (d) show the extent of delamination for each coupon.

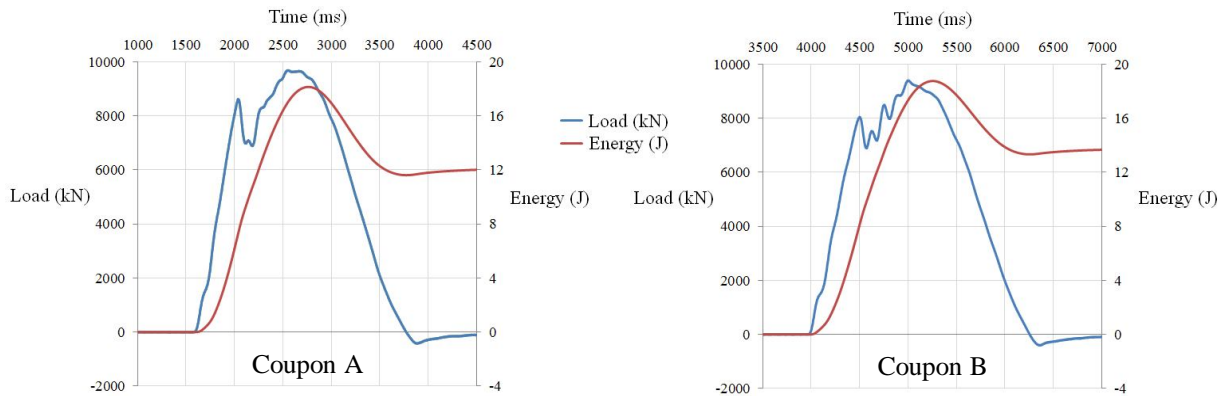

**Figure 5.** Impact load and energy versus time plots.

Highlighted delaminations are those related to sublaminates which subsequently buckle under compressive loading. C-scan images of the impact surfaces of A and B shown in Figs. 6 (a) and (b) indicate that surface damage was of an order of visibility on the boundary of BVID and Clearly Visible Impact Damage (CVID). This is a consequence of the surface ply fibre failure seen as cracking on Figs. 6 (a) and (b). As shown in Figs 6 (c) and (d) the full delamination area for coupon A is larger than for coupon B. Delamination morphologies appear to be similar for both coupons although the presence of a tow gap in coupon B can clearly be seen on Fig. 6 (d).

*4.2 Compression after impact results*

For all DIC images in Fig. 6, axial compressive displacement was applied vertically, colours show out-of-plane displacement from an unloaded reference state and thin well defined contours indicate a steep gradient. Two separate sublaminate buckling events were detected in each coupon test. The sublaminate relating to the smaller $1^{st}$ [45/-45] interface delamination (diagonally orientated red (or mid-grey) delaminations in Figs. 6 (e) and (f)) was the first to buckle in both coupons. In coupon B buckling of this sublaminate was clearly constrained by the tow gap running through the centre of the laminate, see Fig. 6(f). Similarly, the formation of the second sublaminate buckling mode in coupon B (blue (dark grey) delaminations, Fig. 6(d)) was inhibited by the same tow gap, (see Fig. 6 (h)). No interference of sublaminate buckling by tow gaps was seen in coupon A.

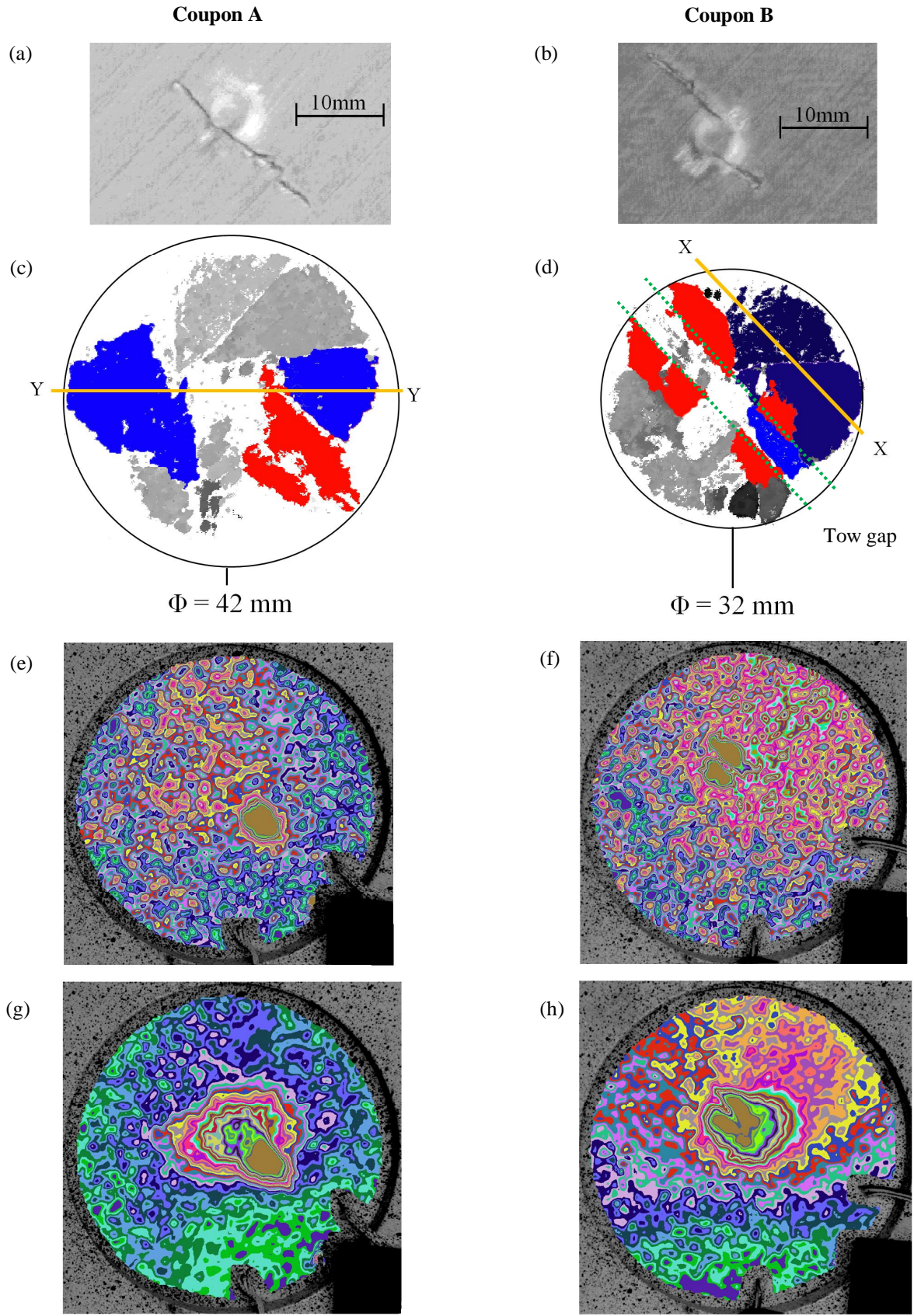

**Figure 6.** Left hand images relate to coupon A and right hand images to coupon B: (a) and (b) surface C-scan images of impact sites. (c) and (d) time-of-flight C-scan images of impact damage viewed from the free surface. Red areas indicate delaminations involved in 1$^{st}$ sublaminate buckling events and blue areas delaminations in 2$^{nd}$ sublaminate buckling events. (e) and (f) DIC images of fully formed 1$^{st}$ ply buckles (at 61 kN and 85 kN respectively). (g) and (h) DIC images of multiple sublaminate buckles immediately prior to propagation at 184kN (335 N/mm$^2$) and 211kN (383 N/mm$^2$) respectively.

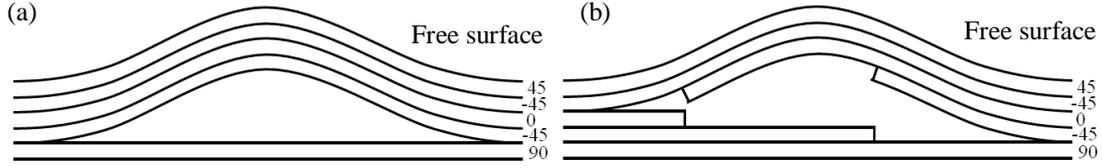

**Figure 7.** (a) and (b) Idealized schematics of sections YY and XX from Figs 6(c) and (d) respectively following the second sublaminate buckling event in each coupon.

Comparison of C-scan and DIC images in Fig. 6 show that the formation of the 2$^{nd}$ sublaminate buckle in coupon B occurs above a number of different delaminations. A similar phenomena has been reported by Greenhalgh et al. (2009). In contrast, a comparison of Figs. 6 (c) and (g) indicates that the second sublaminate buckle in coupon A formed when two or more delaminations at the 4$^{th}$ interface coalesced. Based on sections through XX and YY in Figs. 6 (c) and (d), Fig. 7 shows an idealized cross-sectional representation of the layers involved in the second sublaminate buckling mode for each coupon. Failure occurred in both laminates following the formation of the 2$^{nd}$ sublaminate buckle as a consequence of unstable delamination propagation [Rhead et al. (2011)].

Figure 8 gives experimental and analytical sublaminate buckling and delamination propagation results for both coupons based on delamination diameters derived from Figs. 6(c) and (d). Experimental strains were calculated by correlating average strain gauge readings with DIC images (see Figs 6 (g) and (h)) using the load output of the compression test machine which is captured on the DIC images. The average of the strain gauge readings was used as it accounts for losses in stiffness during the tests.

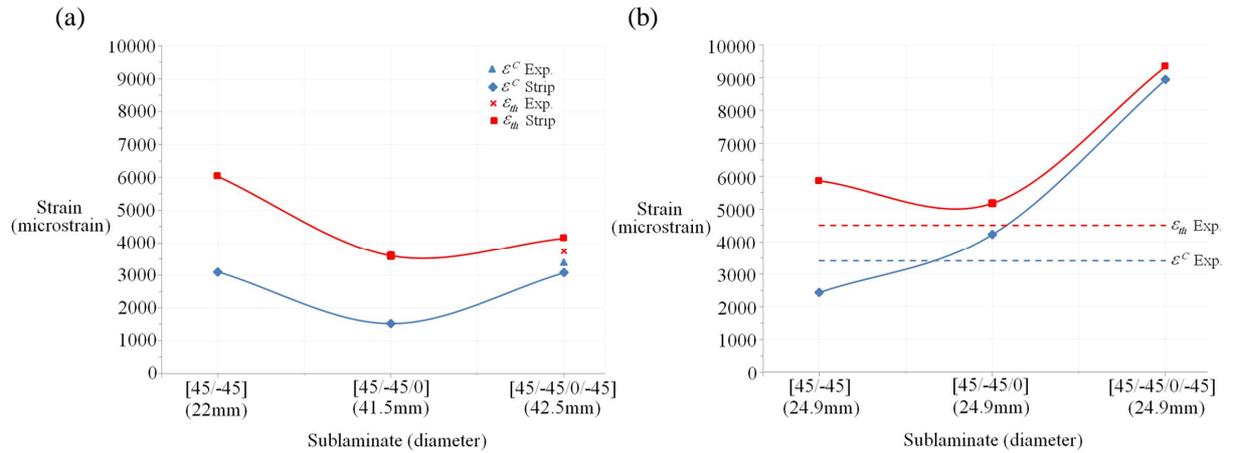

**Figure 8.** Experimental and analytical sublaminate buckling and delamination propagation strains for the outer 2, 3 and 4 ply sublaminates of coupon (a) A and (b) B. Diameters of delamination considered in the analysis are given in brackets.

*4.3 Analytical results*

VICONOPT sublaminate buckling strains in Fig. 8 are based on a circular approximation of individual delamination areas using the greatest extent of the delamination as a diameter. DIC images are used to pinpoint the delaminations involved and are then correlated with C-scans taken prior to compression to accurately determine delamination extent; contrast all images in Fig. 6. For coupon A analytical results are given for 2, 3 and 4-ply sublaminates using the appropriate delamination diameters and associated values of $A_{11}$. However, experimental results are only plotted for the 4-ply sublaminate in Fig. 8(a) as it is clear from Fig. 6 that this was the sublaminate below which critical delamination growth occurred. A comparison of Figs 6(c) and (g) indicates that the 2$^{nd}$ sublaminate buckling event in coupon B involved a sublaminate with areas that were two, three and four plies thick. Hence for coupon B, as a bounding approximation to the actual multi-thickness sublaminate, results are given for sublaminates consisting of 2, 3 and 4 continuous plies with the same delamination diameter, see Fig. 8(b).

# 5 Discussion of tow gap and Strip model results

*5.1 Impact damage*

Results indicate that the position, width and depth of tow gaps have a significant effect on damage resistance. A comparison of C-scans in Figs. 6(c) and (d) shows that the presence of a tow gap near the free surface directly under the point of impact results in a smaller total area of delamination in coupon B than in coupon A. It is suggested that the difference in extent of delaminations between coupons A and B may be a consequence of the tow gap acting as a crack blunter; impeding the spread of delaminations in Mode II (shearing) during impact. Additionally, the tow gap may inhibit through-thickness shearing (Mode III) during impact, particularly near the non-impact face. The marginally increased elastic response to impact noted in coupon B may also, to some extent, account for the smaller delaminations seen in this coupon. However, as coupons were otherwise identical, this elasticity is likely to have been a consequence of the tow gap. It remains to be seen whether this improved resistance will apply to impact in the vicinity of a tow gap.

*5.2 Compression after impact*

A comparison of DIC and C-scan images in Fig. 6 clearly shows that the first sublaminate buckle in both coupons contains only a single ply. The split in the buckling modeshape of the single ply sublaminate in coupon B is due to the interference of the tow gaps in the $1^{st}$ and $8^{th}$ plies from the free surface. The delaminations relating to $1^{st}$ ply sublaminates in both coupons failed to propagate following buckling. This was captured by the Strip model which predicts $1^{st}$ interface delamination growth at over 8000 μstrain for both coupons A and B.

Experimental results in Fig. 8 indicate that failure occurred at a higher applied strain for coupon B than for coupon A indicating that tow gaps near the free surface may be beneficial for damage tolerance. It is suggested that this is either a consequence of the smaller total area of impact damage in coupon B or the result of the tow gap preventing the delaminations at the $4^{th}$ interface from joining up; as occurred in the core of coupon A. Both of these possibilities are linked to the presence of a tow gap near the free surface under the point of impact. It is noted that comparisons could be made with a CAI test on a coupon that is free of tow gaps. However, current industry AFP manufacturing techniques (unlike CTS [Kim et al. (2012)]) mean tow gaps are unavoidable in tapered parts. Hence, rather than their absence, it is the effect on CAI strength of the frequency and areal distribution of these gaps that are the variables that need to be considered if AFP is to be used to make tapered components.

The prediction of threshold propagation strain for the delamination associated with the second sublaminate to buckle in coupon A was within 11% of the experimental value. A comparison of analytical and experimental results in Fig. 9 (b) shows that the strain at which the $2^{nd}$ sublaminate buckling event in coupon B occurred is bounded by analytical predictions for buckling of the $2^{nd}$ and $3^{rd}$ ply continuous sublaminates. Analytical propagation results for the 3 ply sublaminate are within 15% of the experimental result. Delamination propagation is often linked to the presence of $0^{o}$ plies in the associated sublaminate. Hence, interchanging the $0^{o}$ ply for a $45^{o}$ ply in the area where sublaminates buckled may have improved damage tolerance [Rhead et al. (2011)].

# 6 Conclusions

Two coupons containing tow gaps were subject to compression after impact testing. Tow gaps in coupon A were located close to the impact face. Tow gaps in coupon B were located directly under the impact site near the non-impact surface and caused significantly more surface deformation than in coupon A. The near-surface region adjacent to the non-impact face is where delaminations most likely to propagate under compressive loading form during impact. The surprising result was that impact damage was less severe for coupon B. Equally surprising was that damage tolerance was 14% better for coupon B as both delamination formation during impact and delamination growth during compressive loading were inhibited. This is thought to be a consequence of a 3mm wide channel, located directly under the impact site in Coupon B and caused by consolidation over tow gaps in the $1^{st}$ and $8^{th}$ plies during laminate manufacture, preventing two separate delaminations from coalescing during compression loading. The analytical Strip model was able to correctly predict the failure strain of Coupon A. An accurate post-test analysis of coupon B was also possible despite the non-uniform stiffness and thickness of the initial sublaminate. Significantly results suggest that tow gaps may be beneficial for damage tolerance. However, further work is required to ascertain what effect the presence of tow-gaps in the vicinity of an impact (rather than directly below it) will have on damage resistance and damage tolerance. It is noted that it may be possible to derive an optimal distribution of deliberate tow-gaps for improved compression after impact strength.

**Acknowledgements**

The authors thank Dr Richard Newley (GKN Aerospace) for supplying test pieces and for technical advice.